\documentclass[10pt]{article}
\usepackage{mathrsfs}
\usepackage{amsmath,amsfonts,amssymb,mathtools}
\usepackage{cite}
\usepackage{dsfont}
\usepackage{graphicx}
\usepackage{hyperref}
\usepackage{verbatim}
\usepackage{slashed}
\usepackage[all]{xy}
\usepackage{xcolor}
\usepackage{subfig}
\usepackage{tikz}
\usetikzlibrary{shapes.geometric,arrows}
\usepackage{relsize}
\usepackage{caption}
\usepackage{float}
\usetikzlibrary{positioning}
\usepackage{geometry}
\usepackage[T1]{fontenc}
\usepackage{lmodern}
\usepackage{stackengine} 
 
\usepackage[utf8]{inputenc}

\hypersetup{
  colorlinks,
  citecolor=violet,
  linkcolor=blue,
  urlcolor=blue}

\csname @addtoreset\endcsname{equation}{section}
\DeclareMathAlphabet\mathbfcal{OMS}{cmsy}{b}{n}

\newcommand{\beq}{\begin{equation}}
\newcommand{\eeq}{\end{equation}}
\newcommand{\bea}{\begin{eqnarray}}
\newcommand{\eea}{\end{eqnarray}}
\newcommand{\ba}{\begin{array}}
\newcommand{\ea}{\end{array}}
\newcommand{\bit}{\begin{itemize}}
\newcommand{\eit}{\end{itemize}}
\newcommand{\nn}{\nonumber}

\newcommand{\mezzo}{\frac{1}{2}}
\newcommand{\complesso}{{\ \hbox{{\rm I}\kern-.6em\hbox{\bf C}}}}
\newcommand{\reale}{{\hbox{{\rm I}\kern-.2em\hbox{\rm R}}}}
\newcommand{\uno}{ \,  \raisebox{+0.14em}{{\hbox{{\rm \scriptsize ]}} \raisebox{-0.2em}{\kern-.8em\hbox{1}}}} \, }  

\newcommand{\p}{\partial}

\renewcommand{\a}{\alpha}

\newcommand{\D}{\Delta}
\newcommand{\e}{\epsilon}

\renewcommand{\k}{\kappa}

\renewcommand{\L}{\Lambda}

\newcommand{\m}{\mu}

\newcommand{\n}{\nu}
\renewcommand{\r}{\rho}
\newcommand{\s}{\sigma}

\renewcommand{\c}{\chi}

\newcommand{\x}{\xi}

\newcommand{\om}{\omega}

\begin{document}


\begin{titlepage}

\vspace{0.3cm}

\begin{flushright}
$LIFT$--6-1.24
\end{flushright}

\vspace{0.3cm}

\begin{center}
\renewcommand{\thefootnote}{\fnsymbol{footnote}}
\vskip 9mm  
{\huge \bf Most general Type D Black Hole
\vskip 4mm
  and the Accelerating 
\vskip 6mm  
  Reissner-Nordstrom-NUT-(A)dS solution
}
\vskip 33mm
{\large {Marco Astorino$^{a}$\footnote{marco.astorino@gmail.com} 
}}\\

\renewcommand{\thefootnote}{\arabic{footnote}}
\setcounter{footnote}{0}
\vskip 8mm
\vspace{0.2 cm}
{\small \textit{$^{a}$Laboratorio Italiano di Fisica Teorica (LIFT),  \\
Via Archimede 20, I-20129 Milano, Italy}\\
} \vspace{0.2 cm}

\end{center}

\vspace{4.3 cm}

\begin{center}
{\bf Abstract}
\end{center}
{The Plebanski-Demianski metric is commonly thought to describe the most general Type D black hole in general relativity (possibly coupled with Maxwell electromagnetism), but in the form we know at the moment, it fails to include the elusive accelerating spacetimes with gravitomagnetic mass without angular momentum. We propose a very general type D metric, stemming from a binary black hole system, which represents a novel branch of the type D spacetimes: accelerating Kerr-Newman-NUT black hole with cosmological constant. It has straightforward limits to all the known type D black holes in the Einstein-Maxwell theory and in particular to the accelerating Reissner-Nordstrom-NUT solution, yet unknown in the literature.}

\end{titlepage}

\addtocounter{page}{1}

\newpage


\section{Introduction}
\label{sec:introduction}

Analytical and exact black hole solutions are of the utmost importance for understanding the nature of the large scale structure of our universe and for testing challenging fundamental physical concepts both in general relativity and in any other gravitational theory. The most common and studied black hole solutions, such as the Kerr-Newman family, belongs to the type D of the Petrov classification. It is believed that the Debever, Plebanski-Demianski metric \cite{Debever} -\cite{Plebanski-Demianski}, representing an accelerating and dyonically charged rotating black hole endowed with gravitomagnetic parameter and possibly the cosmological constant, is the most general type D black hole solution in four-dimensional General Relativity with standard Maxwell electromagnetism. It is also often believed that the Plebanski-Demianski metric encloses all black hole type solutions of the D class. Nevertheless some issues undermine these beliefs. For example it is not currently known whether or not the Plebanski-Demianski metric might contain all the type D accelerating black hole metrics with the NUT parameter \cite{Podolsky-nut},\cite{New-Look}. This matter is tricky because the Plebanski-Demianski certainly contains a certain specialization of the Type D accelerating Kerr-Newman-NUT metric \cite{Podolsky-New-Lambda}, \cite{PD-NUTs}, however when switching off the angular momentum in that solution also the acceleration disappear. This makes it difficult to retrieve the accelerating NUTty black hole without angular momentum. It is unclear, at the moment, if this issue is related to a perfectible parameterization or a loss of generality of the Plebanski-Demianski metric.  \\
To overcome this impasse recently a large family of type-D accelerating Kerr-NUT black holes has been built, by means of the inverse scattering technique, in \cite{marcoa-equivalence} for the Einstein gravity theory in four-dimensions\footnote{Actually in \cite{marcoa-equivalence} a very general solution for all types of axisymmetric and stationary accelerating black holes in pure general relativity is explicitly written. It describes a single accelerating and rotating black hole or an arbitrary number of accelerating and rotating black holes in a collinear configuration, where each of the event or accelerating horizons brings his own independent set of integration constants associated to mass, angular momentum and NUT parameter. (The mass parameter in for the accelerating horizon is interpreted the acceleration parameter, according with the Einstein equivalence principle.)}. These metrics have been interpreted as a rotating black hole in the near horizon background of a huge Schwarzschild black hole. Specifically it has been shown in \cite{marcoa-equivalence} that this general black hole family, stemming from binary systems, contains also accelerating black holes with gravitomagnetic mass, but also possibly without angular momentum, in particular the (type D) accelerating Taub-NUT metric.\\
In this article we would like to extend the above results in the presence of the Maxwell electromagnetic field, in order to be able to describe also charged black holes, covering the full domain of the Plebanski-Deminaski family. Thus we might be able to model an entire new branch of accelerating black holes containing, for instance, the type D accelerating Reissner-Nordstrom-NUT black hole, which remains unknown in the literature so far.\\
Note that we are focusing only in type D black holes, because a more general class of accelerating and NUTty black holes, such as type-I, has already been constructed in \cite{mann-stelea-chng}, \cite{tesi-giova}, \cite{PD-NUTs} and \cite{Type-I}, basically through symmetries transformation of the Ernst equations. However these metrics do not represent black holes in a standard Rindler background, but the accelerating background carries also extra features such as NUT or electromagnetic fields. Therefore the interpretation of the type I metrics is less intuitive with respect to the type D family. Moreover, when the NUT parameter is not present, these type I spacetimes are plagued by curvature singularity outside the event horizon.\\

\section{Type-D accelerating Kerr-Newman-NUT-$\L$}
\label{Kerr-Newman-NUT-L}

In this article we deal with Einstein general relativity (possibly with cosmological constant) coupled with Maxwell electromagnetism. The field equations for the metric $g_{\m\n}$ and the vector potential $A_\m$ are
\bea  \label{field-eq-g}
                        &&   R_{\m\n} -   \frac{1}{2} R \ g_{\m\n} + \L g_{\m\n} =   2 \left( F_{\m\r}F_\n^{\ \r} - \frac{1}{4} g_{\m\n} F_{\r\s} F^{\r\s}  \right) \ ,   \\
       \label{field-eq-A}                  &&   \nabla_\m F^{\m\n}  = 0  \  .
 \eea
The Maxwell potential and electromagnetic field are related, as usual, by $F_{\m\n}:=\nabla_\m A_\n- \nabla_\n A_\m$.\\
We would like to write the most general type D black hole metric for this theory\footnote{In this paper we consider only type D metrics with two expanding repeated principal null directions, non-expanding type D metrics can be found in \cite{Debever:1983pi}.}. In practice we have to generalise the result of \cite{marcoa-equivalence} to include the electromagnetic monopole charges for the black hole. Thus we aim to remain in the realm of axisymmetric and stationary spacetimes, whose most general metric in four dimensions, possess two commuting Killing vectors, $\p_t$ and $\p_\varphi$ and it can be written in the Lewis-Weyl-Papapetrou, as in (\ref{KNN-inizio}). Using the methods provided in \cite{charging}, i.e. relying only of an analytic behaviour of the electromagnetic Ernst potential with respect to the gravitational one, we find that the metric can be written as\footnote{A Mathematica notebook with the solution and its fundamental limiting spacetimes can be found between the arXiv source files, for the reader convenience.}
\beq \label{KNN-inizio}
        ds^2 = -f(r,x) \left[ dt - \omega(r,x) d\varphi \right]^2 + \frac{1}{f(r,x)} \left[ e^{2\gamma(r,x)}  \left( \frac{{d r}^2}{\Delta_r(r)} + \frac{{d x}^2}{\Delta_x(x)} \right) + \rho^2(r,x) d\varphi^2 \right] \ ,
\eeq
with
\bea
    f&=&  \frac{ \{[1+\a^2(\ell^2-a^2)x^2]^2\D_r-[a+2\a \ell r + a \a^2 r^2]^2\D_x \}}{ (1+\a r x)^2 \ \mathcal{R}(r,x) } \ \ell \ ,  \nn  \\
    \omega &=& \frac{(a-2\ell x + ax^2)[1+\a^2(\ell^2-a^2)x^2]\D_r + (r^2+\ell^2-a^2)(a+2\a \ell r + \a^2 a r^2 )\D_x}{[1+\a^2(\ell^2-a^2)x^2]^2\D_r-[a+2\a \ell r + a \a^2 r^2]^2\D_x} + \omega_0 \ , \nn \\
    \gamma &=& \mezzo \log \left\{C_{f_0} \ \frac{[1+\a^2 x^2(\ell^2-a^2)]^2\D_r-(a+ 2 \a \ell r + a \a^2 r^2)^2\D_x}{ (1+\a r x)^4} \right\}   \ , \nn \\
    \mathcal{R} &=&  \ell\ \big\{ (\ell-ax)^2 +2r\ell \a(a+ax^2-2x\ell) + x^2\a^2(a^2-\ell^2)^2 +r^2[1+\a^2(a-x\ell)^2] \big\} \ ,\nn \\
    \rho &=&\frac{\sqrt{\D_r(r) \ \D_x(x)}}{(1+\a rx)^2} \ , \nn \\
     \D_r &=& (1-\a^2 r^2) [(r-m)^2-\s^2] \  , \nn \\
    \D_x &=& (1-x^2)[(1+\a m x)^2-\a^2x^2\s^2] \ .
     \label{KNN-fine} 
\eea
The constant $C_{f_0}$ is an arbitrary real number undetermined by the field equations, for convenience it can be fixed as $(1+\a^2 a^2)^{-1}$. For the moment we are considering $\L=0$, the generalisation to non-null cosmological constant can be found below.\\
This metric is supported by an electromagnetic field that can be deduced by the following vector potential
\beq
        A_\m (r,x) = \Big[ A_t(r,x) , \ 0 \ , \ 0 \ , \ A_\varphi(r,x) \Big] \ , \nn
\eeq
where
\bea
        A_t  &=& \frac{\sqrt{1-\frac{\ell^2\a^2}{1+a^2 \a^2}}\Big\{r \big(p r - e \ell + a p \ell \a + a^2 p \a^2 r \big) -x\ell(p+ae\a) (a+2r\ell\a+ar^2\a^2)  } {\mathcal{R}(r,x)}  \nn\\
        &+& \frac{x^2(a+r\ell \a)\big[ap+e\ell^2\a + a^3 p \a^2 -a p \ell^2 \a^2 + r\ell\a(p+ae\a)\big] \Big\} } {\mathcal{R}(r,x)} - \frac{p}{\ell}\ ,  \label{KNN-A-fine}   \\
        A_\varphi &=& -\frac{\sqrt{1-\frac{\ell^2\a^2}{1+a^2 \a^2}}\Big\{ar(pr-e\ell  +a p \ell\a + a^2pr\a^2) +x^2(a^2-\ell^2)\Big[e\ell^2\a + a p \big[1+(a^2-\ell^2) \a^2 \big] \Big] }{\mathcal{R}(r,x)} \nn\\
        &+&  \frac{x\ell \big[r(2\ell-xa-xr\ell\a)(e-ap\a) + (\ell^2-a^2)(p+ae\a)+r^2(ae\a-p-2a^2p\a^2) \big]}{\mathcal{R}(r,x)} \Big\}-  \omega_0 A_t + \frac{ap}{\ell} \ . \nn 
\eea

Note that this electromagnetic field has two principal null directions  aligned with the two expanding repeated principal null directions of the Weyl metric tensor. Note also that this metric precisely coincides with the uncharged one presented in \cite{marcoa-equivalence}. The only difference lies in the value of the constant
\beq
         \s = \sqrt{m^2+\ell^2-a^2-e^2-p^2} \ ,
\eeq
which naturally generalises the neutral $\s$ by the presence of the electric and magnetic charges $e$ and $p$. The remaining parameters of the metric, $\a, m , \ell , a$ are related to the acceleration, the mass, the NUT parameter and the angular momentum respectively. The inner, event and acceleration horizons are determined by $\D_r(r)=0$, that is
\beq
          r_\pm = m \pm \s  \ , \hspace{3cm}   r_\a= \frac{1}{\a} \ .  \label{horizons}
\eeq 
The conformal infinity is located, as usual accelerating black holes at $r=1/(\a x)$. The conserved charges exactly coincide with the value of their parameters when the acceleration is null and we remain basically with the Kerr-Newman-NUT black hole. $\omega_0$ is an arbitrary gauge constant related to the angular velocity of the reference frame. It can be set to move the position of the Misner string or to impose a specific asymptotic behaviour of the spacetime.    \\
In the general case the above solution represents an accelerating Kerr-Newman-NUT black hole, therefore at first sight one may think it is contained in the Plebanski-Demianski family, also because both metrics belong to the Petrov type-D class. Nevertheless there are some differences. For instance the black hole characterizations of the Plebanski-Demianski metric are divided in two disjoint sectors: the accelerating black hole branch and the NUTty black hole branch. In fact the Plebanski-Demianski class of black holes is though not to describe an accelerating black holes with gravitomagnetic mass, without angular momentum, such as the NUTty C-metrics \cite{Podolsky-nut}, \cite{Podolsky-New-Lambda}. That's because these metrics should belong to the intersection of the two separated branches. On the contrary the spacetime (\ref{KNN-inizio})-(\ref{KNN-fine}) is explicitly designed to describe the accelerating Taub-NUT spacetimes, as shown in \cite{marcoa-equivalence} for the neutral case.     \\
Thanks to \cite{charging}, it is not difficult to extend that result to the electromagnetically charged case, i.e. to generate the accelerating Reissner-Nordstrom-NUT, as done in below, also in the presence of the cosmological constant.\\
The Petrov D type of the solution (\ref{KNN-inizio})-(\ref{KNN-fine}) have been computed by checking the scalar invariant equality $I^3=27 J^2$, for details see \cite{stephani-big-book} (or \cite{Type-I}).\\
The $\ell$ is related to the presence of the Misner string as can be easily seen by
\beq \label{Dw}
      \D \om =   \lim_{x \to 1}  \om(r,x) - \lim_{x \to -1} \om(r,x) = - \frac{4\ell}{1 - a^2 \a^2 + \ell^2 \a^2 } \ .
\eeq
Thus the only way to remove the Misner string consist in vanishing the NUT parameter: $\ell=0$. \\

\subsection{The inclusion of the cosmological constant}
\label{sec:cosmo}

In the presence of the cosmological constant the above solution can be extended just shifting the functions $\D_r , \D_x$, as in the neutral case \cite{marcoa-equivalence}
\bea
\hspace{-0.6cm}     &&     \D_r(r)  \to  \D_r(r) + \L \left\{ \ell^2 \k + \ell \e r + \left[ a (\a\e +\x) + \a^2 \ell^4 \c - (1+a^2\a^2)(\k-a^2\chi) - \ell^2 \left(\frac{1}{1+a^2\a^2} + 2a^2\a^2\c  \right) \right] r^2 \right. \nn \\
\hspace{-0.6cm}        &&  \hspace{3.3cm} \left. - \ \a \ell r^3\left( \frac{4a}{3+3a^2\a^2} -\x \right) - \left( \frac{1}{3} +\a^2\ell^2 \c \right) r^4  \right\} , \nn \\
\hspace{-0.6cm} &&         \D_x(x)  \to  \D_x(x) + \L \left\{ \ell^2 \c + \ell \x x  + \left(\frac{4a\ell}{3+3a^2 \a^2} +\a \ell \e \right)x^3 - \left( \frac{a^2+(a^2-\ell^2)^2\a^2}{3+3a^2\a^2}  + \ell^2\a^2\k\right) x^4 \right.  \label{cosmo}  \\
\hspace{-0.6cm} &&   \hspace{3.3cm} \left. + \left[\k+a^2\a^2\k-\a^2\ell^4\c + \ell^2\left( 2a^2\a^2 \c - \frac{1}{1+a^2\a^2} \right) -a \left(\x+a\c+\a(\e+a^3\a\c) \right) \right] x^2 \right\}  . \nn
\eea
Of course some of the integrating constants ($\e, \kappa, \xi, \chi$) have to be properly rescaled to obtain the desired limit to the known subcases. If $C_{f_0}$ is not fixed as above the cosmological constant in eqs. (\ref{cosmo}) have to be dilated by a factor $C_{f_0} (1+\a^2a^2)$. \\
Notice that the position of the inner and outer horizon $r_\pm$ is affected by the presence of the cosmological constant, hence differs from values in (\ref{horizons}).\\
The condition (\ref{Dw}) for the absence of the Misner strings in this cosmological case is more involved. In that case $\D\omega$ can be null without vanishing $\ell$, however the constraints on the integrating constants influence drastically the physical interpretation of the metric.\\

\subsection{Further generalisation}
 
While for the black hole description the parametrization of the solution as in eqs. (\ref{KNN-inizio})-(\ref{KNN-fine}) is already optimal, it could be useful to write the complete form of the $\D_r$ and $\D_x$ for a direct connection with the standard Plebanski-Demianski metric, which, whether possible, it is quite non-trivial, so it will be done elsewhere.\\
The most general form of the $\D_r$, $\D_x$ functions, with $\L=0$, compatible with (\ref{KNN-inizio})-(\ref{KNN-fine}) is in fact
\bea \label{Dw-general}
       \D_r(r) &=& (1-\a^2 r^2) [(r-m)^2-\s^2] \ + \ r_0 + r_1 r + r_2 r^2 + r_3 r^3 \a + r_4 r^4 \a^2  \nn  \ , \\
       \D_x(x) &=&(1-x^2)[(1+\a m x)^2-\a^2x^2\s^2] \ - \ \frac{r_4}{\a^2} + \frac{r_3}{\a} x -r_2 x^2 + r_1x^3\a - r_0 x^4\a^2 \ ,
\eea
with $\big\{r_i\big\}_{i=0...4}$  \ real constants and
\beq
        r_0 = \frac{-a^2r_4+ar_3\ell-r_2\ell^2+ar_1\ell\a - r_4 \a^2 (a^2 - \ell^2)^2}{1+\a^2a^2} \ .
\eeq
When the cosmological constant is non null the same extension presented in section \ref{sec:cosmo} can be also applied.\\

\subsection{Type D accelerating Reissner-Nordstrom-NUT-$\L$}
\label{sec:Acc-RNN}

When the angular momentum parameter vanishes, i.e. $a=0$, the solution (\ref{KNN-inizio})-(\ref{KNN-A-fine}) get simplified into
\bea
    f(r,x) &=&  \frac{(1+\a^2 \ell^2 x^2)^2\D_r-4\a^2 \ell^2 r^2\D_x}{(1+\a r x)^2 [(r^2 + \ell^2)(1+\ell^2 \a^2 x^2) - 4rx\a \ell^2]} \ ,  \label{Acc-RNN-inizio}  \\
 \omega(r,x) &=&   \frac{2\ell[\a r(r^2+ \ell^2)\D_x - x(1+\a^2\ell^2x^2)\D_r]}{(1+\a^2 \ell^2 x^2)^2\D_r-4\a^2 \ell^2 r^2\D_x}\ +\omega_0 \ ,  \nn \\
   \gamma(r,x) &=& \mezzo \log \left[\frac{(1+\a^2 \ell^2 x^2)^2\D_r-4\a^2 \ell^2 r^2\D_x}{(1+\a r x)^4} \right]   \ ,  \nn \\
  A_t(r,x)  &=& -\frac{p}{\ell} +\frac{r \sqrt{1-\ell^2\a^2} [pr-e\ell-2 x p \ell^2 \a + x^2 \ell^2(pr+e\ell)\a^2]}{\ell \ [(r^2 + \ell^2)(1+\ell^2 \a^2 x^2)-4rx\a \ell^2]} \ , \nn \\
  A_\varphi (r,x) &=& \frac{ x \sqrt{1-\ell^2\a^2} [p(r^2-\ell^2)+e\ell(r^2x\a-2r+x\ell^2\a)]}{(r^2 + \ell^2)(1+\ell^2 \a^2 x^2)-4rx\a \ell^2} - \om_0 A_t(r,x) \ . \label{Acc-RNN-fine}
\eea
This solution, unknown to the literature so far, probably because it is not clear if it can be derived from the Plebanski-Demianski family, represents the (type-D) accelerating Reissner-Nordstrom-NUT-(A)dS spacetime, as can be easily understood by its limits for null NUT parameter, null electromagnetic charges or null acceleration. Indeed, from (\ref{Acc-RNN-inizio}) -(\ref{Acc-RNN-fine}), when the NUT parameter vanishes, and the angular velocity of the asymptotic observer is fixed to zero by setting $\om_0=0$, we get the accelerating Reissner-Nordstrom-(A)dS black hole:
\bea
      ds^2  &=&  \frac{1}{(1+\a r x)^2} \left[ - Q(r) dt^2 - \frac{dr^2}{Q(r)} + \frac{r^2 dx^2}{P(x)} + r^2 P(x) d\varphi^2 \right] \ , \\
       A  &=&  - \frac{e}{r} \ dt \  + \  p x \ d \varphi  \ ,  
\eea
where
\bea
      Q(r) &=&  \left(1-\frac{2m}{r} + \frac{e^2+p^2}{r^2} \right) \left(1-\a^2 r^2\right) - \frac{\L}{3} \left(r^2 +3\k \right) \ , \\
      P(x) &=& (1-x^2)\big\{1+\a x \big[2m+(e^2+p^2)x\a \big]\big\} + x^2\L \k \ .
\eea
For the usual accelerating Reissner-Nordstrom-(A)dS black hole interpretation usually the parameter $\k$ is thought to be null.\\

While when the accelerating parameter is null we obtain, from  (\ref{Acc-RNN-inizio}) - (\ref{Acc-RNN-fine}), the Reissner-Nordstrom-NUT-(A)dS solution. However before taking the $\a \to 0$ limit it is better to rescale the integrating constants as follows: $\k \to \k / \ell^2 , \chi \to \chi / \ell^2 , \xi \to \xi/\ell , \epsilon \to \epsilon/\ell$,  then we get
\bea
      ds^2  &=&   - \frac{F(r)}{r^2+\ell^2} \big[dt-(\om_0-2x\ell) d\varphi\big]^2 + \frac{r^2+\ell^2}{F(r)} \ dr^2 + \frac{r^2+\ell^2}{H(x)} + (r^2+\ell^2) H(x) \ d\varphi^2  \\
       A  &=&  - \frac{er+p\ell}{r^2+\ell^2} \ dt  \ + \ \frac{p x r^2+er(\om_0-2x\ell)+p\ell(\om_0-x\ell)}{r^2+\ell^2} \ d\varphi  \ ,
\eea
where
\bea
        F(r)  &=&  (r^2-2mr - \ell^2 + e^2 + p^2) + \L \left[ \k + \e r - r^2 \left( \ell^2 + \frac{\k}{\ell^2}  \right) -\frac{r^4}{3}  \right]  \ , \\
        H(x)  &=& (1-x^2) + \L \left[ \chi + x  \xi - x^2 \left ( \ell^2 - \frac{\k}{\ell^2} \right) \right]  \ .
\eea
Notice that usually $\kappa = \epsilon = \xi = 0 , \chi =\ell^2 , \omega_0 =2\ell $ to have the standard form of the Reissner-Nordstrom-NUT-(A)dS. \\
When the electromagnetic charges are switched off we get the (type-D) accelerating Taub-NUT-(A)dS explicitly written in \cite{marcoa-equivalence}.

These accelerating black hole metrics endowed with gravitomagnetic mass are not to be confused with their type I counterparts, typically generated, thanks to the Ehlers or the Harrison transformation of the Ernst equations \cite{enhanced}. We are referring to the type I accelerating Taub-NUT, found in \cite{mann-stelea-chng} (see also \cite{Podolsky-nut}), or the type I accelerating Reissner-Nordstrom-NUT and the type I accelerating Kerr-Newman-NUTs presented respectively in \cite{tesi-giova} and \cite{PD-NUTs} (for further generalisations see \cite{Type-I}). Apart from manifest geometrical differences encoded in their different Petrov type, the type I and the type D accelerating NUTty metrics are shown in \cite{Type-I} and \cite{marcoa-equivalence} to have a determined physical differences. In fact the type D accelerating black hole metrics can be thought of as a black hole close to the event horizon of a huge Schwarzschild black hole. On the other hand the accelerating type I black holes stems from binary systems where the bigger black hole carries extra features such as electromagnetic, gravitomagnetic charges or angular momentum. In any case both types of solutions can be obtained as limits of binary systems, where one of the black holes grows until it becomes the accelerating horizon of the other black hole, for details see \cite{marcoa-equivalence} and \cite{Type-I}. \\

\begin{figure}[H]

\tikzstyle{rect} = [draw,rectangle, fill=white!20, text width =3cm, text centered, minimum height = 1.5cm,scale=0.82]

\begin{tikzpicture}

\useasboundingbox (-8.1,-15) rectangle (6,1.5);

\node[rect,scale=1.2,anchor=south,label=above:{(\ref{KNN-inizio})-(\ref{KNN-A-fine})+(\ref{cosmo})},text width=4cm,line width=1.2pt](Full){\bf Accelerating Kerr-Newman-NUT-(A)dS \hspace{0.1cm}\\ ($\alpha, a, e, p, \ell, m, \Lambda $)};
\node[rect,scale=1.2,anchor=north,label=above:{\hspace{-1.3cm} (\ref{Acc-RNN-inizio})-(\ref{Acc-RNN-fine})+(\ref{cosmo})},text width=4.2cm,line width=1.2pt,below of=Full,node distance=5cm](acc kerr newman nut){\bf Accelerating Reissner-Nordstrom-NUT-(A)dS \\ ($\alpha, e, p, \ell, m, \Lambda $)};  
\node[rect,scale=1.2,anchor=north, label=above left:{\cite{marcoa-equivalence}},text width=3.6cm,line width=1.2pt,left of=acc kerr newman nut,node distance=5cm](acc kerr newman nut uno){\bf Accelerating Kerr-NUT-(A)dS \\ ($\alpha, a, \ell, m, \Lambda $)}; 
\node[rect,scale=1.2,anchor=north,label=above:{\hspace{2.2cm}(\ref{KNN-inizio})-(\ref{KNN-A-fine})},text width=3.6cm,line width=1.2pt,right of=acc kerr newman nut, node distance=5cm](acc kerr newman nut due){\bf Accelerating Kerr-Newman-NUT ($\alpha, a, e, p, \ell, m$)}; 
\node[rect,scale=1.2,anchor=north, label=above left:{\cite{marcoa-equivalence}},text width=3.6cm,line width=1.2pt,below of=acc kerr newman nut uno, node distance=3cm](acc kerr newman nut sei){\bf Accelerating Kerr-NUT  \\ ($\alpha, a, \ell, m$)};
\node[rect,scale=1.2,anchor=north,label=above:{\hspace{2.4cm} (\ref{Acc-RNN-inizio})-(\ref{Acc-RNN-fine})},text width=4cm,line width=1.2pt,below of=acc kerr newman nut due, node distance=3cm](acc kerr newman nut sette){\bf Accelerating Reissner-Nordstrom-NUT \\ ($\alpha, e, p, \ell, m$)}; 
\node[rect,scale=1.2,anchor=north, label=above:{\hspace{-1.8cm}\cite{marcoa-equivalence}},text width=3.6cm,line width=1.2pt,below of=acc kerr newman nut, node distance=3cm](acc kerr newman nut otto){\bf Accelerating Taub-NUT-(A)dS \\ ($\alpha, \ell, m, \Lambda$)}; 
\node[rect,scale=1.2,anchor=north,text width=3.6cm,below of=acc kerr newman nut sei, node distance=3cm](acc kerr newman nut nove){Accelerating Kerr \\  ($\alpha, a, m$)};
\node[rect,scale=1.2,anchor=north,text width=3.6cm, below of=acc kerr newman nut sette, node distance=3cm](acc kerr newman nut dieci){Accelerating Reissner-Nordstrom \\ ($\alpha, e, p, m $)}; 
\node[rect,scale=1.2,anchor=north, label=above:{\hspace{-1cm}\cite{marcoa-equivalence}},text width=3.6cm,line width=1.2pt,below of=acc kerr newman nut otto, node distance=3cm](acc kerr newman nut undici){\bf Accelerating Taub-NUT \\ ($\alpha, \ell, m$)}; 
\node[rect,scale=1.2,anchor=north,text width=3cm,below of=acc kerr newman nut nove, node distance=3cm](acc kerr newman nut dodici){Rindler\\  ($\alpha$)};
\node[rect,scale=1.2,anchor=north,text width=3.4cm, below of=acc kerr newman nut dieci, node distance=3cm](acc kerr newman nut tredici){Reissner-Nordstrom \\ ($e, p, m $)}; 
\node[rect,scale=1.2,anchor=north,text width=3.6cm,below of=acc kerr newman nut undici, node distance=3cm](acc kerr newman nut quattordici){ Accelerating Schwarzschild \\ ($\alpha, m$)}; 
\node[rect,scale=1.2,anchor=north,text width=3.3cm,above left of=acc kerr newman nut uno, node distance=3.8cm](acc kerr newman nut pippo){Kerr-Newman-NUT-(A)dS \\ ($a, e, p, \ell, m, \Lambda $)}; 
\node[rect,scale=1.2,anchor=north,text width=3.4cm,above right of=acc kerr newman nut due, node distance=3.8cm](acc kerr newman nut pippa){Accelerating Kerr-Newman-(A)dS ($\alpha, a, e, p, m, \Lambda $)}; 

\draw[
->] (Full) -- node [right,near start] {\; \; $\ell$ = 0}(acc kerr newman nut pippa);
\draw[
->] (Full) -- node [left,near start] {$\a$ = 0 \; \;}(acc kerr newman nut pippo);
\draw[->] (Full) -- node [left] {$e$ = $p$ = 0} (acc kerr newman nut uno);
\draw[->] (Full) -- node [left] {$a$ = 0} (acc kerr newman nut);
\draw[->] (Full) -- node [right] {\; $\L$ = 0} (acc kerr newman nut due);
\draw[->] (acc kerr newman nut due) -- node [left, near start] {$a$ = 0} (acc kerr newman nut sette);
\draw[->] (acc kerr newman nut uno) -- node [left] {$\L$ = 0} (acc kerr newman nut sei);
\draw[->] (acc kerr newman nut uno) -- node [left] {$a$ = 0 \;} (acc kerr newman nut otto);
\draw[->] (acc kerr newman nut) -- node [right] {$e=p$ = 0 \;} (acc kerr newman nut otto);
\draw[->] (acc kerr newman nut otto) -- node [right] {$\L$ = 0 \;} (acc kerr newman nut undici);
\draw[->] (acc kerr newman nut) -- node [right] {\; $\L$ = 0} (acc kerr newman nut sette);
\draw[->] (acc kerr newman nut undici) -- node [left] {$\ell$ = 0 \;} (acc kerr newman nut quattordici);
\draw[->] (acc kerr newman nut quattordici) -- node [above] {$m$ = 0 \;} (acc kerr newman nut dodici);
\draw[->] (acc kerr newman nut nove) -- node [left] {$a$ = 0 \;} (acc kerr newman nut quattordici);
\draw[->] (acc kerr newman nut sei) -- node [left] {$\ell$ = 0 \;} (acc kerr newman nut nove);
\draw[->] (acc kerr newman nut sei) -- node [left] {$a$ = 0 \;} (acc kerr newman nut undici);
\draw[->] (acc kerr newman nut sette) -- node [right] {$\ell$ = 0 \;} (acc kerr newman nut dieci);
\draw[->] (acc kerr newman nut sette) -- node [right] {\; $e=p$ = 0 \;} (acc kerr newman nut undici);
\draw[->] (acc kerr newman nut dieci) -- node [right] {\; $e=p$ = 0\;} (acc kerr newman nut quattordici);
\draw[->] (acc kerr newman nut dieci) -- node [right] {$\a$ = 0\;} (acc kerr newman nut tredici);

\end{tikzpicture}


\caption{Structure of the new general class of type-D black hole solutions (\ref{KNN-inizio})-(\ref{KNN-A-fine})+(\ref{cosmo}) and all its specializations. The spacetimes not included in the parametrization of the Plebanski-Demianski family (known so far) are emphasized in bold characters. Essentially the novel metrics are all the ones that contain simultaneously the accelerating and NUT parameter. While apparently a sort of type-D accelerating Kerr-Newman-NUT-(A)dS spacetimes can be included in a parametrization of the Plebanski-Demianski they behave differently with respect to (\ref{KNN-inizio})-(\ref{KNN-A-fine})+(\ref{cosmo}), because even in the best Plebanski-Demianski black hole parametrization available at the moment \cite{New-Look}, \cite{Podolsky-New-Lambda}, it is not known how to switch-off the angular momentum without vanishing also the acceleration. On the contrary (\ref{KNN-inizio})-(\ref{KNN-A-fine})+(\ref{cosmo}) contains all the limiting subcases well defined, including the type-D accelerating Reissner-Nordstrom-NUT-(A)dS and accelerating Taub-NUT-(A)dS specializations.\\}
\label{fig:albero}
\end{figure}
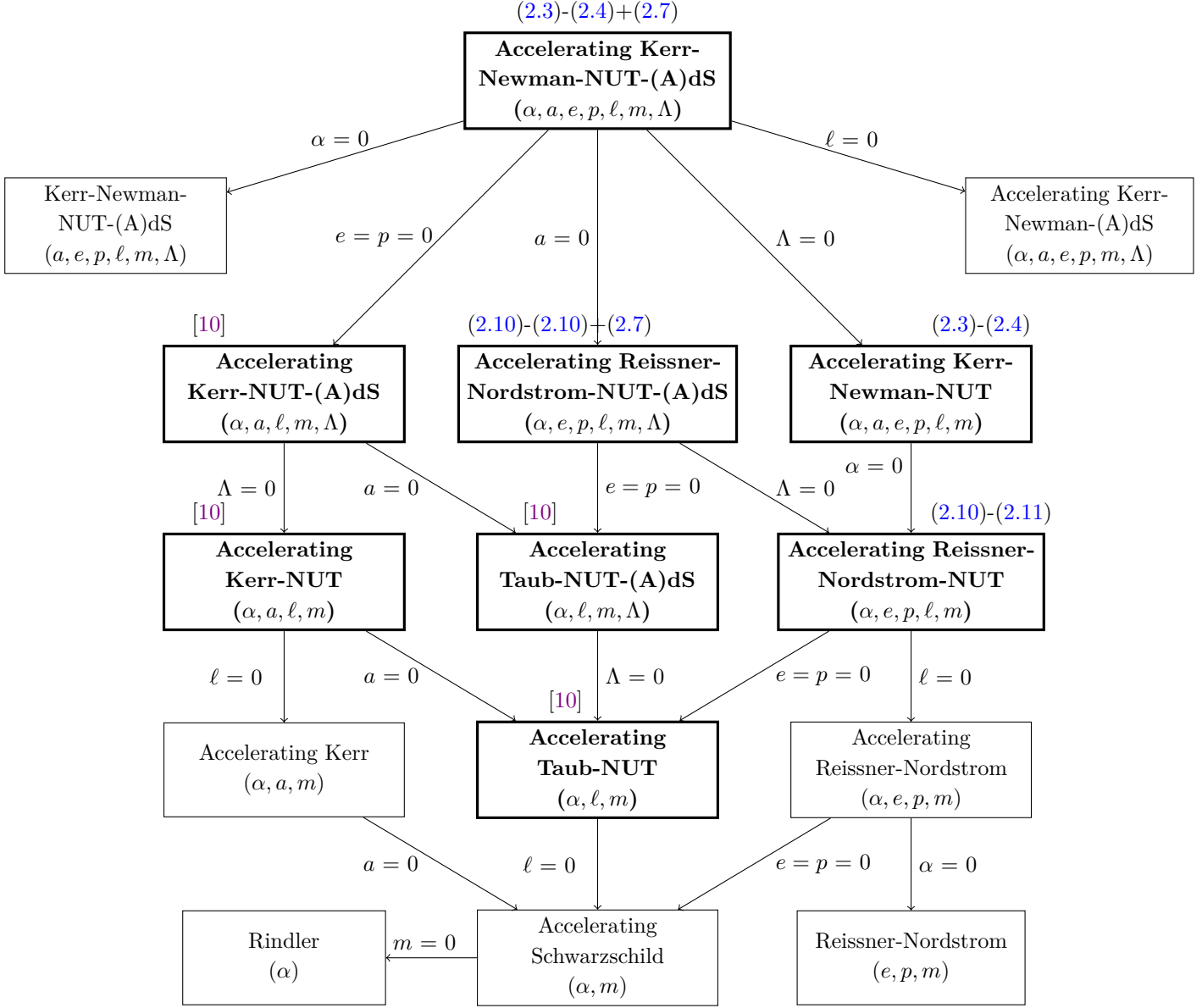

In figure \ref{fig:albero} the main specializations of the solution (\ref{KNN-inizio})-(\ref{KNN-A-fine})+(\ref{cosmo}) are portrayed.

\section{Conclusions}

In this article we propose a new general solution to describe all the black hole family of Einstein general relativity, belonging to the type D class of the Petrov classification, eventually coupled with standard Maxwell electromagnetism and also possibly coupled with the cosmological constant. \\
The main advantages of this type D metric with respect to the previous solutions known so far in the literature are that: $(i)$ all the limit to the special cases are easily accessible and well defined and $(ii)$ accelerating black holes with NUT parameter, but without angular momentum, clearly can be derived from the main metric. \\
This new metric is presented in a parametrization, induced by the inverse scattering technique, specially suitable for the black hole description. Remarkably it is able to describe, for the first time, a new branch of spacetimes including the elusive type D accelerating Taub-NUT and accelerating Reissner-Nordstrom-NUT black hole, which in the previous forms of the type D metrics remained hidden.\\
As it has been shown for the uncharged case, in \cite{marcoa-equivalence}, the best physical interpretation of these new solutions comes from a binary black hole limit: These spacetimes describe a charged and NUTty black hole close by a huge Schwarzschild black hole, whose event horizon become, in the near horizon limit of the uncharged black hole, a Rindler horizon. In this picture the acceleration is not generated by the pulling string but, thanks to the Einstein equivalence principle, by the presence of the gravitational field of the big Schwarzschild black hole. The string tension prevents the collapse while maintaining an equilibrium configuration. This dual interpretation physically explains why the conical singularity cannot be removed from these spacetimes: because two black holes tend to collapse into each other if not sustained. Moreover this interpretation has a conceptual advantage: when the small black hole charges (mass, electromagnetic charges, angular momentum, ...) vanish, also the string vanishes and we naturally remain only with the near horizon metric of the huge Schwarzschild black hole, that is the Rindler horizon. On the other hand, in this no black hole (and no conical singularity) limit, it is not clear why, according to the standard picture,  we should remain with a Rindler horizon, since there is no string providing acceleration.  \\
Because the Debever and the Plebanski-Demianski metrics \cite{Debever} - \cite{Plebanski-Demianski} should describe all the type D accelerating black holes, we suspect that also this new solution can be possibly re-conducted to the Plebanski-Demianski form, but this step is not straightforward. If possible it will be pursued elsewhere \cite{OPA}. Eventually the correspondence with the usual form of the type D solution can be useful to directly extend our result in case of different theories of gravity, such as, for instance, the scalar tensor ones. \\  
Notice also that the novel form of the accelerating Kerr-Newman-NUT black hole presented here is not necessarily physically equivalent to the previous known variants \cite{New-Look},\cite{Podolsky-New-Lambda}. In fact the interpretation of the metric parameters is surely different and the causal structure of the spacetime does not necessarily coincide, even if the Killing horizons are formally in the same location. For instance our new parametrization implies that the conformal infinity is not located as in \cite{Podolsky-New-Lambda}.\\

\paragraph{Acknowledgements}
{\small We would like to thank the Universidad Austral de Chile (UACh) in Valdivia for the hospitality while part of this work was done. We thank Jiri Podolsky for valuable discussions. A Mathematica notebook containing the main solutions presented in this article can be found in the arXiv source folder. 
}\\



\end{document}